\documentclass[12pt]{article}
\usepackage{amsmath, amssymb, amsthm, enumerate, bbm, fullpage, cite}
\usepackage[all,cmtip]{xy}

\newcommand{\CY}{\chi}
\newcommand{\Galt}{\tilde{G}}

\begin{document}
\rightline{ZMP-HH/15-3} 

\vskip .4 true cm  
\begin{center}  
{\Large 10D to 4D Euclidean Supergravity over a Calabi-Yau three-fold}\\[.5em] 
\vskip 0.3 true cm  

{Wafic A.\ Sabra$^{1}$ and Owen Vaughan$^{2}$} \\[.6em] 
$^1${Centre for Advanced Mathematical Sciences and Physics Department\\ 
American University of Beirut\\ 
College Hall,
P.O.Box: 11-0236, Beirut, Lebanon\\  
ws00@aub.edu.lb}\\  
$^2${Department of Mathematics and Center for Mathematical Physics\\ 
Universit\"at Hamburg\\ 
Bundesstra{\ss}e 55, 
D-20146 Hamburg, Germany\\  
owen.vaughan@math.uni-hamburg.de}\\

March 17, 2015, revised May 20, 2015, revised \today

\end{center}

\begin{abstract}  

\noindent  
We dimensionally reduce the bosonic sector of 10D Euclidean type IIA supergravity over a Calabi-Yau three-fold. The resulting theory describes the bosonic sector of 4D, ${\cal N} = 2$ Euclidean supergravity coupled to vector- and hyper-multiplets.

We show that the scalar target manifold of the vector-multiplets is projective special para-K\"ahler, and is therefore of split signature, whereas the target manifold of the hyper-multiplets is (positive-definite) quaternionic K\"ahler.

\end{abstract}  

\vspace{-1.5em}
\tableofcontents

\section{Introduction}

Supersymmetric Euclidean theories coupled to vector-multiplets have recently
been a subject of interest \cite{mohaupt1, mohaupt2, mohaupt3}. It has been
known for some time that the complex scalar fields of vector-multiplets
in 4D, ${\cal N}=2$ Lorentzian supersymmetric theories exhibit so-called
special K\"{a}hler geometry \cite{witvan84}. This geometry has provided a
useful tool in the understanding of field theory non-perturbative structure,
supergravity, string compactifications (see for example \cite{onebigref}),
as well as in the study and analysis of black hole physics \cite{blackdev}.

Both rigid and local Euclidean versions of special geometry  were
constructed and analysed in terms of para-complex geometry in \cite{mohaupt1,
mohaupt2, mohaupt3}. Roughly speaking, the Euclidean versions of special
geometry can be obtained from the standard versions appearing in Lorentzian theories by
replacing $i$ with the para-complex unit $e$, 
which satisfies the properties $e^{2}=1$ and $\bar{e} =-e $. 
In the supergravity literature such a replacement first appeared in
the study of D-instantons in type IIB supergravity \cite{gb}%
\footnote{In this reference the para-complex unit $e$ is referred to as the hyperbolic complex unit.}. 
Para-complex manifolds are necessarily of even dimension and split signature.
For further details on para-complex geometry and Euclidean supersymmetric theories we refer the reader to \cite{mohaupt1}.
It is important to emphasise that it is only the target geometry of the scalar fields
that becomes para-complex in Euclidean theories. This is not true of the superalgebra representation itself or the geometry of superspace, which are both complex in the case of Euclidean spacetime signature.

Throughout this paper we will use the convention that the degree of supersymmetry ${\cal N}$ of a Euclidean superalgebra is matched to the number of real supercharges in the Lorentzian case. For example, 4D, ${\cal N} = 2$ Euclidean supersymmetry has 8 real supercharges. This is despite the fact that the smallest supersymmetry representation in 4D Euclidean space has 8 real degrees of freedom \cite{Zumino:1977yh}, and therefore there is no 4D, ${\cal N}$ = 1 Euclidean theory in our conventions.

The rigid 4D, ${\cal N}= 2$ Euclidean vector-multiplet action 
was constructed in \cite{mohaupt1} by reducing 5D, ${\cal N} = 2$
vector-multiplets over a time-like circle.
The Euclidean action and supersymmetry
transformation rules were expressed in terms of para-holomorphic
coordinates. 
Similarly in the local case, 
the bosonic sector of 4D, ${\cal N}=2$ Euclidean supergravity 
coupled to vector-multiplets was constructed in \cite{mohaupt3} by reducing 5D, ${\cal N} = 2$ 
supergravity coupled to vector-multiplets \cite{GST}
over a timelike circle, 
and the scalar target manifold was shown to be 
projective special para-K\"{a}hler \cite{mohaupt3}.
The Killing spinor equations as well as 
the classification of supersymmetric
gravitational instanton solutions of these theories were later analysed in 
\cite{inst1,inst2}. 

Theories of ${\cal N} = 2$ hyper-multiplets are also of interest.
The scalar target manifold in 3, 4 and 5 dimensions is hyper-K\"ahler in the rigid case \cite{AlvarezGaume:1981hm} 
and quaternionic K\"ahler in the local case \cite{Bagger:1983tt}. Since the hyper-multiplet target manifold is invariant under dimensional reduction, this indicates that the target manifold of 4D local Euclidean hyper-multiplets is also quaternionic K\"ahler.
On the other hand, 4D vector-multiplets (both rigid and local) can be mapped to 3D hyper-multiplets by dimensional reduction followed by
Hodge dualisation, which is known as the supergravity $c$-map. Reducing over a spacelike circle results in a theory of 3D hyper-multiplets with Lorentzian spacetime signature and quaternionic K\"ahler target manifold \cite{Ferrara:1989ik}.
However, reducing over a timelike circle results in a theory of 3D hyper-multiplets with Euclidean spacetime signature and para-quaternionic K\"ahler target manifold \cite{Vaughan:2012, Cortes:2015}. 
This suggests that, at least in certain circumstances, the target geometry of Euclidean hyper-multiplets is not completely fixed by the signature of spacetime. Therefore, one must be careful to identify the correct target geometry for the theory in question.

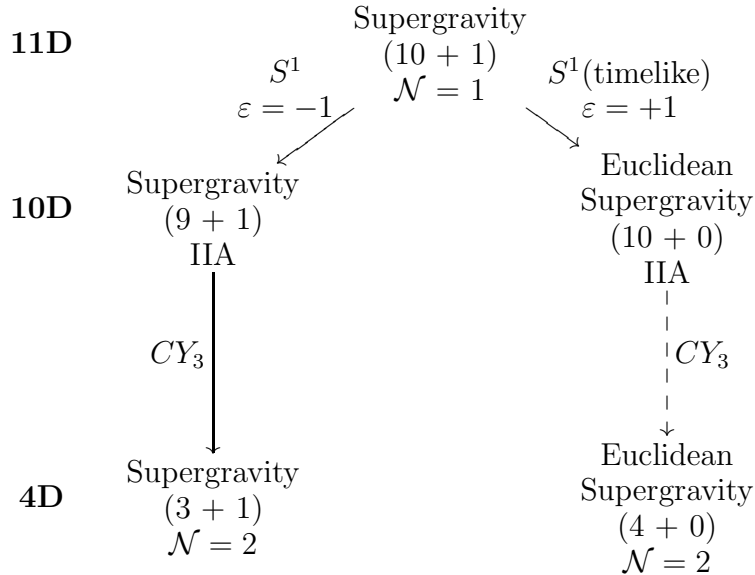
\begin{figure}[h]
\[
	\xymatrix@R=5mm@C=5mm
	{
		\textbf{11D}
		& 
		& \txt{Supergravity \\ (10 + 1) \\ ${\cal N}$ = 1 } \ar[dl]_{\txt{$S^1$ \\ $\varepsilon = -1$}} \ar[dr]^{\; \; \txt{$S^1$(timelike) \\ $\varepsilon = +1$}}
		& 
		& \hspace{1.5em} \\
		\textbf{10D} 
		& \txt{Supergravity \\ (9 + 1) \\ IIA }   \ar[ddd]_{\txt{$CY_3$}}
		& 	
		& \txt{Euclidean \\ Supergravity \\ (10 + 0) \\ IIA } \ar@{-->}[ddd]^{\txt{$CY_3$}} 
		& \\
		\\
		\\
		\textbf{4D}
		& \txt{Supergravity \\ (3 + 1) \\ ${\cal N} = 2$ }    
		& 
		& \txt{Euclidean \\ Supergravity \\ (4 + 0) \\  ${\cal N} = 2$ }   
		&
	}
\]
\caption{\emph{The dimensional reduction of supergravity from \emph{11D} to \emph{4D} over a spacelike or timelike circle and Calabi-Yau three-fold. In this paper we complete the second step in the right hand side of this diagram.}}
\label{fig:1}
\end{figure}

The goal of this paper is to establish the higher-dimensional origins of 
4D, ${\cal N} = 2$ Euclidean supergravity,
and, in particular, how the scalar target geometry emerges through the process of dimensional reduction. 
Our starting point is standard 11D supergravity with Lorentzian spacetime signature.
We first reduce this theory over a timelike circle in order to obtain 10D Euclidean type IIA supergravity \cite{Hull}, and then reduce this 10D theory further over a Calabi-Yau three-fold. 
Whilst the dimensional reduction of 11D supergravity over tori with one timelike circle has been considered in \cite{Hull:1998br, Cremmer:1998em}, which in our context would correspond to taking the Calabi-Yau manifold to be $T^6$, 
 the reduction over a timelike circle followed by an arbitrary Calabi-Yau three-fold is currently missing from the literature.
The resulting effective theory describes the bosonic sector of 4D, ${\cal N} = 2$ Euclidean supergravity coupled to $h_{1,1}$ vector-multiplets and $(h_{2,1} + 1)$ hyper-multiplets.  Our construction is summarised in figure \ref{fig:1}.

We will follow closely the original work of \cite{Bodner} in which the reduction of 11D supergravity over a spacelike circle followed by a Calabi-Yau three-fold was first constructed.
Indeed, we find that the resulting action of the 4D Euclidean
theory differs from the Lorentzian case only by certain sign flips. We will
keep track of these signs using the parameter $\varepsilon $, which is
determined by the reduction of 11D supergravity over either a spacelike or
timelike $S^1$ according to the rule 
\begin{equation}
\varepsilon =%
\begin{cases}
-1\;, & S^1 \ \text{spacelike} \\ 
+1\;, & S^1 \ \text{timelike}\;.%
\end{cases}
\label{eq:Epsilon}
\end{equation}%
Therefore, after reducing further over a Calabi-Yau three-fold we end up with 4D, 
${\cal N} = 2$ supergravity if $\varepsilon =-1$ and 4D, ${\cal N} = 2$
Euclidean supergravity if $\varepsilon =+1$. We will show that the target space geometry of the 4D scalar fields is given by the product $M_{vector} \times M_{hyper}$, where $M_{vector}$ is a $2h_{1,1}$-dimensional projective special $\varepsilon$-K\"ahler manifold and $M_{hyper}$ is a $(4h_{2,1} + 4)$-dimensional quaternionic K\"ahler manifold.

The (pseudo-)Riemannian structure of spacetime in our construction is given as follows. The spacetime manifolds in various dimensions are related topologically by
\begin{equation}
	M_{11} = S^1 \times M_{10}  \;, 
	\qquad
	M_{10} = \chi  \times M_4\;,
	\label{manifolds}
\end{equation}
where $\chi$ is a Calabi-Yau three-fold. 
The spacetime metrics are related by
\begin{equation}
	g_{11} = -\varepsilon e^{\frac43 \phi'} (d \tilde{x}^0 + V)^2 + e^{-\frac16 \phi'} g_{10} \;,
	\qquad
	g_{10} = g_\chi  +  {\cal V}^{-1} g_4\;,
	\label{metrics}
\end{equation}
where $\phi'$ and $V$ are the 11D Kaluza-Klein scalar and vector respectively, $\tilde{x}^0$ is the coordinate of the $S^1$ dimension, ${\cal V}$ is the volume of the Calabi-Yau three-fold and $g_\chi$ is the Ricci-flat Calabi-Yau metric. The signatures of the various spacetime metrics are 
\begin{align*}
	\text{sig}( g_{11} ) = (-\varepsilon,\varepsilon,+,\ldots,+) \;,
	\qquad
	\text{sig}( g_{10} ) = (\varepsilon,+,\ldots,+) \;,
	\qquad
	\text{sig}( g_{4} ) = (\varepsilon,+,+,+) \;.
\end{align*}
The internal compact (pseudo-)Riemannian manifold $S^1 \times \chi$ has metric $g_{S^1 \times \chi} =  -\varepsilon e^{\frac43 \phi'} (d \tilde{x}^0 )^2 + e^{-\frac16 \phi'} g_{\chi}$ that depends on the base point in $M_4$. It has signature $(-\varepsilon,+,\ldots,+)$.

\section{10D Euclidean supergravity}

Our starting point is the bosonic part of the 11D supergravity action \cite%
{cremmer} 
\begin{equation}
S^{11}=\int_{M_{11}} \left[  \frac{1}{2}{}^{\ast }R_{11}-\frac{1}{2}\tilde{F}_{4}\wedge
{}^{\ast }\tilde{F}_{4}-\frac{\sqrt{2}}{6}\tilde{F}_{4}\wedge \tilde{F}%
_{4}\wedge \tilde{A}_{3} \right]\;, \notag % 
\end{equation}%
which has spacetime signature $(-\varepsilon ,\varepsilon ,+,\ldots ,+)$ in
coordinates $(\tilde{x}^{0},\ldots ,\tilde{x}^{10})$. We will reduce this
theory over the $\tilde{x}^{0}$ dimension, which we assume is either a spacelike or
timelike circle according to the rule \eqref{eq:Epsilon}. The 11D spacetime manifold and metric decompose into their 10D counterparts according to  \eqref{manifolds} and \eqref{metrics}. The three-form is decomposes according to 
\begin{equation}
\tilde{A}_{3}={A}_{3}^{\prime }+d\tilde{x}^{0}\wedge B_{2}\;,\qquad \tilde{F}_{4} ={F}_{4}^{\prime
}-d\tilde{x}^{0}\wedge H_{3}\;, \notag
\end{equation}%
where ${A}_{3}^{\prime }$ and $B_2$ are degenerate and invariant along the  $\tilde{x}^{0}$
direction and $F'_4 = dA'_3, \, H_3 = dB_2$. This resulting 10D action is given by 
\begin{align}
S^{10}& =2\pi \rho \int_{M_{10}} \Bigg[ \frac{1}{2}{}^{\ast }R_{10}-\frac{1}{4}d\phi
^{\prime} \wedge {}^{ \ast }d\phi ^{\prime }+\frac{1}{4}\varepsilon \,e^{\frac{3}{2}\phi
^{\prime }}dV\wedge {}^{\ast }dV+\frac{1}{2}\varepsilon \,e^{-\phi ^{\prime
}}H_{3}\wedge {}^{\ast }H_{3}  \notag \\
& \hspace{4em}-\frac{1}{2}e^{\frac{1}{2}\phi ^{\prime }}\left( {F}%
_{4}^{\prime }+V\wedge {H}_{3}\right) \wedge {}^{\ast }\left( {F}%
_{4}^{\prime }+V\wedge {H}_{3}\right) -\frac{1}{\sqrt{2}}\left( {F}%
_{4}^{\prime }\wedge {F}_{4}^{\prime }\wedge B_{2}\right) \Bigg] \;, \notag
\end{align}%
which has spacetime signature $(\varepsilon ,+,\ldots ,+)$. Here $\rho $ is
the radius of the $\tilde{x}^{0}$ dimension which we now set to $%
\rho =\frac{1}{2\pi }$. It is convenient to make the field redefinitions 
\begin{equation}
{A}_{3}^{\prime }=A_{3}+V\wedge B_{2}\;,\qquad {F}_{4}^{\prime
}=F_{4}+dV\wedge B_{2}-V\wedge H_{3}\;,\qquad \phi ^{\prime }=\frac{3}{2}%
\log \phi \;, \notag
\end{equation}%
in which case the action becomes 
\begin{align}
S^{10}& =\int_{M_{10}} \Bigg[\frac{1}{2}{}^{\ast }R_{10}-\frac{9}{16}d\log \phi
\wedge {}^{\ast }d\log \phi +\frac{1}{4}\varepsilon \,\phi ^{\frac{9}{4}%
}dV\wedge {}^{\ast }dV  \notag \\
& \hspace{4em}+\frac{1}{2}\varepsilon \,\phi ^{-\frac{3}{2}}H_{3}\wedge
{}^{\ast }H_{3}-\frac{1}{2}\phi ^{\frac{3}{4}}\left( {F}_{4}+dV\wedge {B}%
_{2}\right) \wedge {}^{\ast }\left( {F}_{4}+dV\wedge {B}_{2}\right)  \notag
\\
& \hspace{6em}-\frac{\sqrt{2}}{2}\left( {F}_{4}+dV\wedge B_{2}\right) \wedge 
{F}_{4}\wedge B_{2}-\frac{\sqrt{2}}{6}dV\wedge B_{2}\wedge dV\wedge
B_{2}\wedge B_{2}\Bigg]\;.  \label{10Dact}
\end{align}%
Note that the topological terms in the last line pick up a factor of $\varepsilon$
when written in components, see equation \eqref{eq:10Dlast} in appendix A.

For $\varepsilon =-1$ the action \eqref{10Dact} agrees with the bosonic sector of 10D type IIA supergravity \cite{Bodner}. 
(Note that the final term is not present in \cite{Bodner}. However, it is present in the earlier work \cite{Ferrara:1988ff}.) 
For $\varepsilon = +1$ it agrees with the bosonic sector of 10D type IIA Euclidean supergravity \cite{Hull}. The complete Euclidean supergravity action, including fermionic terms and supersymmetry transformation rules, can be found in \cite{Bergshoeff:2007cg}. 
One can understand the action \eqref{10Dact}  as the field 
theory limit of type IIA string theory with Lorentzian or Euclidean spacetime signature \cite{Hull}.

We would like to dimensionally reduce this theory over a compact six-dimensional internal manifold whilst preserving supersymmetry.  Regardless of the choice of $\varepsilon$, this can be achieved if and only if there exists
a spinor $\eta$ on the internal manifold such that the corresponding infinitesimal supersymmetry transformation of the gravitino vanishes $0 = \delta \Psi = \displaystyle{\not} D \eta$, i.e.\ the internal manifold admits a covariantly constant spinor. 
This motivates us to consider reduction over Calabi-Yau manifolds even in the case of Euclidean spacetime signatures.

\section{Calabi-Yau reduction}

In this section we present some background material which can be found
in \cite{Bodner, Ferrara:1990dp, Candelas}. 

We assume that the 10D spacetime manifold decomposes into $%
M_{10}=\CY \times M_{4}$, where $\CY$ is a Calabi-Yau three-fold and $%
M_{4}$ is a four-dimensional (pseudo-)Riemannian manifold with metric 
signature $(\varepsilon, +, +, +)$.
On $M_{10}$ one may introduce local coordinates 
\begin{equation}
w^{\hat{\mu}}\;,\;\;\;\;\hat{\mu}=1,\ldots ,10\;, \notag
\end{equation}%
which decompose into coordinates 
\begin{align}
& x^{\mu }\;,\;\;\;\;\mu =1,\ldots ,4\;,  \notag \\
& y^{a}\;,\;\;\;\;a=1,\ldots ,6\;, \notag
\end{align}%
on $M_{4}$ and $\CY$ respectively. It is useful to introduce complex
coordinates on $\CY$ as follows$:$ 
\begin{equation*}
\xi _{1}=\frac{1}{\sqrt{2}}(y_{1}+iy_{2})\;,\qquad \xi _{2}=\frac{1}{\sqrt{2}%
}(y_{3}+iy_{4})\;,\qquad \xi _{3}=\frac{1}{\sqrt{2}}(y_{5}+iy_{6})\;.
\end{equation*}%
In these conventions the volume form satisfies $d^{6}y=id^{3}\xi d^{3}\bar{\xi}=:id^{6}\xi$ and the Hodge duals of $(3,0)$-forms and $(2,1)$-forms satisfy
\begin{equation}
	{}^* \rho_{(3,0)} = -i{\rho_{(3,0)}} \;,
	\qquad
	{}^* \sigma_{(2,1)} = i{\sigma}_{(2,1)} \;.
	\label{eq:degree}
\end{equation}
The inner product of two $(p,q)$-forms is defined by
\[
	\left(\alpha_{(p,q)}, {\beta}_{(p,q)}\right) = \int_\CY \alpha_{(p,q)} \wedge {}^* {\beta}_{(p,q)} \;.
\]

\subsection{Harmonic forms and integrals over a $CY_3$}

\label{sec:Harmonic}

On a Calabi-Yau three-fold there are non-trivial harmonic forms in the $(1,1),(2,1),(3,0)$ and $(3,3)$ cohomology sectors (and their Hodge duals). We use the following basis 
\begin{align}
& (1,1)\qquad V^{A}=V_{i\bar{j}}^{A}d\xi ^{i}\wedge d\bar{\xi}^{\bar{j}}\;,
& A& =1,\ldots ,h_{1,1}  \notag \\
& (2,1)\qquad \Phi _{\alpha}=\frac{1}{2}\Phi _{\alpha ij\bar{k}}d\xi ^{i}\wedge d\xi
^{j}\wedge d\bar{\xi}^{\bar{k}}\;, & \alpha& =1,\ldots ,h_{2,1}  \notag \\
& (3,0)\qquad \Omega =\frac{1}{3!}\Omega _{ijk}d\xi ^{i}\wedge d\xi
^{j}\wedge d\xi ^{k}\;, & &  \notag \\
& (3,3)\qquad v=\frac{1}{3!}J\wedge J\wedge J\;, & & \notag
\end{align}%
where $V^A$ are real. The K\"{a}hler form is given by $J=ig_{i\bar{j}}d\xi ^{i}\wedge d%
\bar{\xi}^{\bar{j}}=M^{A} V^{A}$, where $M^A(x)$ are real scalar fields, and the volume by 
\begin{equation}
\mathcal{V}=\int_\CY v=\int_\CY \sqrt{g}\,d^{6}y= \int_\CY i \sqrt{g}\,d^{6}\xi \;. \notag
\end{equation}

Let us first consider certain integrals relevant for the $H^{2}$ cohomology
sector. Following \cite{Bodner} we define
\begin{align}
\mathcal{K}& =\int_\CY J\wedge J\wedge J\;, & \mathcal{K}_{AB}& =\int_\CY V^{A}\wedge
V^{B}\wedge J\;,  \notag \\
\mathcal{K}_{A}& =\int_\CY V^{A}\wedge J\wedge J\;, & \mathcal{K}_{ABC}& =\int_\CY
V^{A}\wedge V^{B}\wedge V^{C}\;, \notag
\end{align}%
which satisfy $\mathcal{K}=\mathcal{K}_{ABC}M^{A}M^{B}M^{C}$ and $\mathcal{V}%
=\frac{1}{6}\mathcal{K}$. 
A useful formula for any real $(1,1)$-form is given by \cite{Candelas} 
\begin{equation}
{}^{\ast }V^{B}=  -J\wedge V^{B} + \frac{3}{2 \mathcal{K}}J\wedge J  \left(  \int_\CY V^{B} \wedge J \wedge J \right) \;, \notag
\end{equation}%
from which it follows that 
\begin{equation}
G_{AB}(M) :=\frac{1}{2{\cal V}}{\int_\CY V^{A}\wedge {}^{\ast }V^{B}}%
=-3\left( \frac{\mathcal{K}_{AB}}{\mathcal{K}}-\frac{3}{2}\frac{\mathcal{K}%
_{A}\mathcal{K}_{B}}{\mathcal{K}^{2}}\right) \;.   \notag
\end{equation}
In components we have the formulae
\begin{align}
	2{\cal V} \, G_{AB} &= \int_\CY d^6 y \sqrt{g}\left[  V_{i\bar{j}}^A V^{Bi\bar{j}} \right] \;, \label{eq:Comp1} \\
	{\cal K}_{AB} &= \int_\CY d^6 y \sqrt{g} \left[ V_{i\bar{j}}^A V^{Bi\bar{j}} - V_{i\bar{j}}^A V^B_{k\bar{l}} g^{i\bar{j}} g^{k\bar{l}} \right] \;. \label{eq:Comp2}
\end{align}

Let us now turn to the $H^3$ cohomology sector. We will follow the conventions for 
special K\"ahler geometry given in appendix A.
Since the $H^{3}$ sector contains contributions
from $h_{2,1}$ harmonic $(2,1)$-forms and one harmonic $(3,0)$-form indices run from 
$I,J=0,\ldots ,h_{2,1}$. Consider a real cohomology basis 
 $\alpha _{I},\beta ^{I}$ of $H^3$ that satisfies 
\begin{equation}
\int_\CY \alpha _{I}\wedge \beta ^{J}= \delta _{I}^{J}\;.
 \label{eq:RealCo}
\end{equation}%
In the above basis the holomorphic three-form $\Omega 
$ can be written as 
\begin{equation}
\Omega =X^{I}\alpha_{I} -F_{I}\beta^{I}\;, \label{eq:Omega}
\end{equation}%
where $F_I = F_I(X)$ and $X^I$ depends only on the 4D spacetime coordinates $x^\mu$.
Integrating gives 
\begin{equation}
\left(\Omega ,\bar{\Omega}\right)  = \int_\CY i\Omega \wedge \bar{\Omega}=- i\left( X^{I}%
\bar{F}_{I}-F_{I}\bar{X}^{I}\right) =  ||\Omega||^2 {\cal V} \;,
 \notag
\end{equation}%
where   $||\Omega ||^2 = \frac{1}{3!} \Omega_{ijk} \bar{\Omega}^{ijk}$.
The function $||\Omega ||^2$ is in fact completely independent of the coordinates $y^a$ and
depends only on the spacetime coordinates $x^\mu$ \cite[Thm 4.3.2]{Fre:1995bc}.
The derivative  $\Omega _{I}= \frac{\partial }{\partial X^{I}}\Omega$ takes the form \cite{Candelas} 
\begin{equation}
\Omega _{I} = \Phi _{I}+K_{I}\Omega \;, \notag
\end{equation}%
where we have introduced an additional harmonic $(2,1)$-form $\Phi _{0}$ 
that is a linear combination of $\Phi_{\alpha}$, and is defined by the above equation.
This implies that 
\[
	\int_\CY \Omega \wedge \Omega_I = 0 \;,
\]
and therefore $F_I = \frac{\partial}{\partial X^I} F$, where $F=F(X)$ is a
homogeneous function of degree two. Using the homogeneity of $F$ we have
\[
	\left(\Omega ,\bar{\Omega}\right) =- N_{IJ} \bar{X}^I X^J \;,
\]
and therefore $N_{IJ} \bar{X}^I X^J$ is strictly negative due to the positivity of the inner product $(\cdot,{\cdot})$.
The formula for $K_I$ is given by
\begin{equation}
K_{I}=\frac{(N\bar{X})_{I}}{\bar{X}NX}
=-\frac{\partial }{\partial X^{I}}K\;, \notag
\end{equation}%
where $K:=-\log (- \bar{X}NX)=-\log (\Omega ,\bar{\Omega})$, and we are using the notation $\bar{X}NX = N_{IJ} \bar{X}^I X^J$ and $(NX)_I = N_{IJ} X^J$. 
From the fact that $\Omega _{I}
=\alpha _{I}-F_{IJ}\beta ^{J} $ one obtains the following 
expressions for $\alpha _{I},\beta ^{I}$ and $\Phi _{I}:$
\begin{align}
\alpha _{I}& =\Omega _{I}+iF_{IJ}N^{JK}\left( \Omega _{K}-\bar{\Omega}%
_{K}\right)   \label{eq:alpha} \\
\beta ^{I}& =iN^{IJ}(\Omega _{J}-\bar{\Omega}_{J})  \label{eq:beta} \\
\Phi _{I}& =\left( \delta _{I}^{J}-K_{I}X^{J}\right) \left( \alpha _{J}-{F}%
_{JK}\beta ^{K}\right) \;,  \label{Phidecomp}
\end{align}%
which allows one to calculate
\begin{align}
\left(\Phi _{I},\bar{\Phi}_{\bar{J}}\right)&= \int_\CY -i\Phi _{I}\wedge \bar{\Phi}_{\bar{J}} 
= \int_\CY d^6y \sqrt{g} \left[  \frac12 \Phi_{I ij\bar{k}} \bar{\Phi}_{\bar{J}}^{\;\;\;ij\bar{k}}\right] \notag \\
&= \left( N_{IJ}-\frac{(N\bar{X})_{I}(N{X})_{J}}{\bar{X}NX}\right)
= - \mathcal{M}_{I\bar{J}}\;.  \notag
\end{align}%
All other integrals vanish $(\Phi _{I},\Omega )=(\Phi _{I},\bar{\Omega}%
)=(\Phi _{I},\Phi _{J})=(\Omega ,\Omega )=0$.

We will also consider the $(0,2)$-forms $b_\alpha$ defined by
\begin{equation}
b_\alpha = \frac12 {b}_{\alpha \bar{i} \bar{j}} d\bar{\xi}^i \wedge d\bar{\xi}^j 
:= -\frac{i}{2} \frac{1}{||\Omega ||^{2}}\bar{\Omega}_{\bar{i}}^{\;\;kl}
{\Phi}_{\alpha kl \bar{j}} d\bar{\xi}^i \wedge d\bar{\xi}^j \;.  \notag%
\end{equation}
Integrating gives
\begin{equation}
	\left(b_\alpha, \bar{b}_{\bar{\beta}} \right) = \int_\CY d^6 y \sqrt{g} \left[ \frac12  b_{\alpha \bar{i}\bar{j}} \bar{b}_{\bar{\beta}}^{\;\;\;\bar{i}\bar{j}}\right] 
	= \frac{1}{||\Omega^2 ||}  \left( \Phi_\alpha, \bar{\Phi}_{\bar{\beta}} \right)
	=: {\cal V} \Galt_{\alpha {\bar{\beta}}} \;,  \label{hypermetric}
\end{equation}
where we have used $\Omega_{ijm} \bar{\Omega}^{klm} = \epsilon_{ijm} {\epsilon}^{klm}||\Omega||^2 = 2\delta_{[i}^{[k} \delta_{j]}^{l]}||\Omega||^2$
and the fact that $||\Omega^2||$ is independent of the coordinates $y^a$.
The matrix $\Galt_{\alpha \bar{\beta}}$ defines a hermitian metric for the $h_{2,1}$ complex variables $z^\alpha = X^\alpha / X^0$. Due to homogeneity we have $\Galt_{\alpha\bar{\beta}}(X^0,X^1,\ldots,X^{h_{2,1}}) = \Galt_{\alpha\bar{\beta}}(1,z^\alpha,\ldots,z^{h_{2,1}})$.
This metric is projective special K\"ahler with K\"ahler potential $K$ defined previously. The holomorphic prepotential on the corresponding conic affine special K\"ahler manifold is given by $F(X)$. For further details on special K\"ahler geometry we refer the reader to \cite{mohaupt3}.

\subsection{Zero modes}

We begin by considering the zero modes of the 10D metric, which we decompose
according to 
\begin{equation}
(g_{10})_{ \mu \nu }(w)=(g'_4)_{\mu \nu }(x)\;,\qquad (g_{10})_{\mu a}(w)=0\;,\qquad
(g_{10})_{ab}(w)=(g_{\CY})_{ab}(x,y)\;. \notag
\end{equation}%
Note that the components $(g_{10})_{\mu a}$ must vanish since they correspond
to a Killing vector on the Calabi-Yau three-fold, and such continuous isometries 
are incompatible with $SU(3)$ holonomy. See, for example, \cite{Green:1987mn}. 
We have denote the 4D metric with a prime in anticipation of a Weyl transformation that will be made later in \eqref{Weyl}.
Zero modes of the wave operator correspond to
deformations of the Ricci-flat Calabi-Yau metric that preserve the $SU(3)$ structure. These
are given by 
\begin{equation}
i\delta g_{i\bar{j}}=\sum_{A}^{h_{1,1}}\delta M^{A}V_{i\bar{j}}^{A}\;,
\qquad
\delta g_{ij}=\sum_{\alpha }^{h_{2,1}}\delta \bar{z}^{\bar{\alpha} }\bar{b}%
_{\bar{\alpha} ij}\;, \notag
\end{equation}%
with $M^{A}$ and $z^{\alpha}$ defined in the previous section.
Since the Calabi-Yau metric is Ricci flat we have 
\begin{align}
R_{ij} &= R_{i\bar{j}}=R_{\bar{\imath}\bar{j}}=0,  \notag \\
R_{10}(w) &= R_{4}'(x)+g'_4{}^{\mu \nu }\left( R_{\;\mu i\nu }^{i}(x,y)+R_{\;\mu \bar{%
\imath}\nu }^{\bar{\imath}}(x,y)\right) \;. \notag
\end{align}%
The Ricci scalar is explicitly calculated to be
\begin{equation}
\frac12 R_{10}= \frac12R_{4}' -\frac{1}{2}\partial _{\mu }z^{\alpha }\partial ^{\mu }\bar{z}%
^{\bar{\beta} }b_{\alpha \bar{i}\bar{j}}{\bar{b}_{{\bar \beta}}^{\;\;\;\bar{i}\bar{j}}}+\partial _{\mu }M^{A}\partial ^{\mu }M^{B}\left( \frac{3}{2}%
V_{i\bar{j}}^{A}V^{Bi\bar{j}}-V_{i\bar{j}}^{A}V_{l\bar{k}}^{B}g^{i\bar{j}%
}g^{l\bar{k}}\right).
\label{exp}
\end{equation}%
We refer the reader to \cite{Bodner} for details concerning this calculation.

Let us now consider the zero modes of the other bosonic fields. Recall that there 
are no harmonic one-forms on a Calabi-Yau manifold. The dilaton and
Kaluza-Klein vector zero modes are given simply by 
\begin{equation}
\phi (w)=\phi (x)\;,\qquad V=V_{\mu }(x)dx^{\mu }. \notag
\end{equation}%
The zero modes of the two-form $B_{2}$ and three-form $A_{3}$ are given by 
\begin{align}
B_{2}(w)& =\mathcal{B}_{2}(x)+a^{A}(x)V^{A}(y)\;, & & a^{A}\in \mathcal{C}%
^{\infty }(M_{4}),\;\;\mathcal{B}_{2}\in \Omega ^{2}(M_{4}),  \notag \\
A_{3}(w)& =\mathcal{A}_{3}(x)+\mathcal{A}^{A}(x)\wedge V^{A}(y)+\check{A}%
(x,y)\;, & & \mathcal{A}^{A}\in \Omega ^{1}(M_{4}),\;\;\mathcal{A}_{3}\in
\Omega ^{3}(M_{4}), \label{eq:10D4Dforms}
\end{align}%
where 
\begin{equation}
\check{A}(x,y)= 2^{\frac{1}{4}} \zeta ^{I}(x)\alpha _{I}(x,y)+ 2^{\frac{1}{4}}\tilde{\zeta}_{I}(x)\beta ^{I}(x,y) \;,\qquad \zeta ^{I},\tilde{\zeta}%
_{I}\in \mathcal{C}^{\infty }(M_{4})\;.  \label{eq:_Acheck}
\end{equation}%
Recall that harmonic forms on manifold with positive definite metric are
always closed, and, hence, exterior derivatives are given by
$dB_2 = d{\cal B}_2 + da^A\wedge V^A$ etc. 

It will be useful later to write the exterior derivative of $\check{A}$ in the basis $\Phi_I,\bar{\Phi}_I, \Omega, \bar{\Omega}$. This can be achieved by first taking the exterior derivative
\[
	d\check{A} = 2^{\frac{1}{4}}d\zeta^I \wedge \alpha_I +  2^{\frac{1}{4}} d\tilde{\zeta}_I \wedge \beta^I \;,
\]
and then expanding $\alpha_I,\beta^I$ in terms of $\Phi_I,\Omega$ and their complex conjugates%
\footnote{In a previous version of this paper an alternative calculation of $d\check{A}$ was presented, which is included in appendix B. We thank one of our referees for suggesting the more concise calculation presented here.
}.
Using expressions \eqref{eq:alpha}, \eqref{eq:beta} and the fact that $N_{IJ} = i(\bar{F}_{IJ} - F_{IJ})$ one finds, after some simplifications, that
\begin{align*}
	d\check{A} &= i 2^{\frac{1}{4}} \left(d\tilde{\zeta}_J  + \bar{F}_{JK} d\zeta^K \right) N^{JI} \wedge \Phi_I - i 2^{\frac{1}{4}} \frac{1}{(\bar{X}NX)} X^I \left(d\tilde{\zeta}_I  + {F}_{IJ} d\zeta^J \right) \wedge \bar{\Omega} + h.c. \;.
\end{align*} 
Next, observe that
\[
	\bar{F}_{JK} N^{JI} \Phi_I = {\cal N}_{JK} N^{JI} \Phi_I \;,
	\qquad
	X^I {F}_{IJ} = X^I {\cal N}_{IJ} \;,
\]
where in the first equation we used $X^I \Phi_I = 0$ which can easily be seen from \eqref{Phidecomp}. We may now write the derivative as
\begin{equation}
d\check{A}=P^{I}\wedge \Phi _{I}+\bar{Q}\wedge \bar{\Omega}+h.c.\ \;,
\label{F4decomp}
\end{equation}
where $P^{I},\bar{Q}\in \Omega ^{1}(M_{4})\otimes \mathbbm{C}$ are given by 
\begin{align}
P^{I}& =i2^{\frac{1}{4}}\left(d\tilde{\zeta}_{J} + \mathcal{N}_{JK}d\zeta ^{K} \right) N^{JI}\;,  \qquad
\bar{Q} =-i2^{\frac{1}{4}}\frac{1}{(\bar{X}NX)}X^{I}\left(d\tilde{\zeta}_{I} + \mathcal{N}_{IJ}d\zeta ^{J} \right)\;. \notag
\end{align}

Having written down all zero modes our task is to 
construct the corresponding four-dimensional effective action. For reduction
over a Calabi-Yau three-fold it is known that this can be obtain by substituting
the above expressions into the 10D action and integrating over the Calabi-Yau
three-fold. We will show that, as one would expect, one 
obtains a theory of ${\cal N} = 2$ Euclidean supergravity
coupled to $h_{1,1}$ vector-multiplets and $(h_{2,1} + 1)$ hyper-multiplets.  

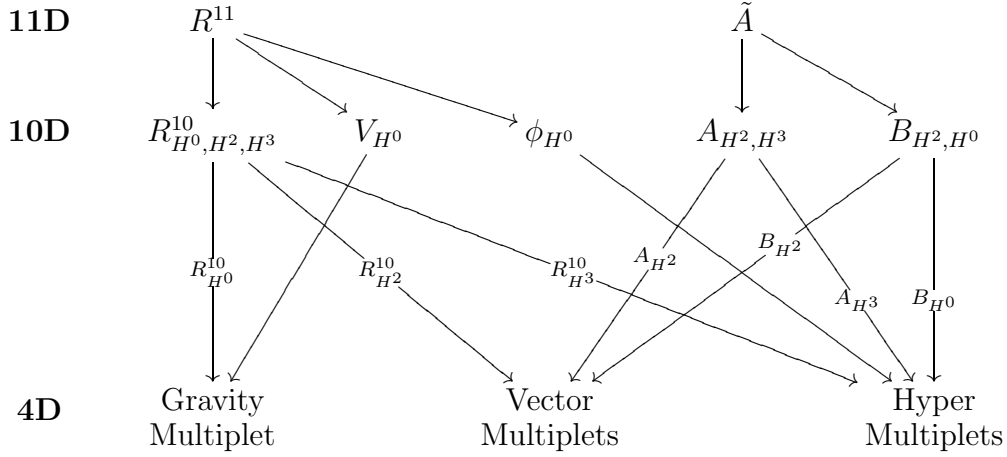
\begin{figure}
\[
	\xymatrix@R=8mm@C=8mm
	{
		\textbf{11D} 
		& {R}^{11} \ar[d] \ar[dr] \ar[drr]
		&
		&
		& \tilde{A} \ar[d] \ar[dr] \\
		\textbf{10D} 
		& R^{10}_{H^0,H^2,H^3} \ar[ddd]|{R^{10}_{H^0}} \ar[dddrr]|{R^{10}_{H^2}} \ar[dddrrrr]|{R^{10}_{H^3}}   
		& V_{H^0} \ar[dddl] 
		& \phi_{H^0} \ar[dddrr] 
		& A_{H^2,H^3} \ar[dddr]|(0.6){A_{H^3}} \ar[dddl]|(0.45){A_{H^2}} 
		& B_{H^2,H^0} \ar[dddll]|(0.4){B_{H^2}} \ar[ddd]|(0.6){B_{H^0}}  \\
		\\
		\\
		\textbf{4D} & \txt{Gravity \\ Multiplet } & & \txt{Vector \\Multiplets } & & \txt{Hyper \\ Multiplets } 
	}
\]
\caption{\emph{Composition of gravity-, vector- and hyper-multiplets.}}
\label{fig:Comp}
\end{figure}

From the four-dimensional perspective we expect to see the 
following field content: 
\begin{equation}
H^{0}\text{-sector}:\;\;\phi ,V,\mathcal{B}_{2} \qquad
H^{2}\text{-sector}:\;\;a^{A},\mathcal{A}^{A} \qquad H^{3}\text{-sector}%
:\;\;\zeta ^{I},\tilde{\zeta}_{I}\;. \notag
\end{equation}%
The one-forms will only appear in the action through their field strengths
${\cal F}^0 = dV$ and ${\cal F}^A = d{\cal A}^A$, which 
form the gauge-fields of the gravity-multiplet and vector-multiplets 
respectively. The two form also only appears
through its field strength ${\cal H}_3 = d{\cal B}_2$, which is then dualised
to a scalar field  $\tilde{\phi}$. This contributes to the 
hyper-multiplet sector along with $\phi$ and $\zeta^I,\tilde{\zeta}_I$.
The composition of gravity-, vector- and hyper-multiplets is displayed
schematically in figure \ref{fig:Comp}.

% Schematic diagram

We end this section by considering the contribution of the $H^{0}$ sector
of $A_{3}$ to the four-dimensional action. After performing the Weyl rescaling 
\eqref{Weyl} this term is given by
\begin{equation}
S_{H^{0}(A_{3})}^{4}=\int_{M_{4}} \Bigg[ -\frac{1}{2} \phi ^{\frac{3}{4}}\mathcal{V}^3(%
\mathcal{F}_{4}+dV\wedge \mathcal{B}_{2})\wedge {}^{\ast }(\mathcal{F}%
_{4}+dV\wedge \mathcal{B}_{2}) \Bigg]\;. \label{eq:Cos}
\end{equation}%
By adding an appropriate Lagrange multiplier the four-form ${\cal F}_4$ may 
be dualised to a constant $e_0$, which appears as a prefactor in front of a  scalar potential of the form $(\phi^{\frac{3}{4}}{\cal V}^3)^{-1}$ in the resulting action.
In order to avoid such a potential we will set $e_0 = 0$, after
which \eqref{eq:Cos} vanishes completely and plays no further role in 
the discussion. 
We remark that in the case of Lorentz spacetime signature this term corresponds to an RR-flux,
 and induces a gauging of the axionic scalar field $\tilde{\phi}$ dual to ${\cal B}_2$ (which will be introduced later in section \ref{sec:axion}) with charge $e_0$ and gauge field $V$ \cite{Louis:2002ny}%
\footnote{We would like to thank one of our referees for pointing this out.}% 
.

\section{4D gravity- and vector-multiplets}

In this section we will consider the contributions from the gravity and $H^2$ cohomology sector to the reduction of the 10D action \eqref{10Dact} over a Calabi-Yau three-fold.
The resulting 4D effective action with spacetime signature $(\varepsilon,+,+,+)$ is given by
\begin{align}
& S_{grav+vector}^{4}=\int_{M_{4}}\Bigg[\frac{1}{2}{}^{\ast
}R_{4}-g_{AB}(y)\left( dx^{A}\wedge {}^{\ast }dx^{B}-\varepsilon
\,dy^{A}\wedge {}^{\ast }dy^{B}\right)   \notag \\
& +\varepsilon \,\Bigg(\frac{1}{2}({c}yyy)\left( \frac{1}{6}+%
\frac{2}{3}(gxx)\right) \mathcal{F}^{0}\wedge {}^{\ast }\mathcal{F}^{0}-%
\frac{2}{3}({c}yyy)(gx)_{A}\mathcal{F}^{A}\wedge {}^{\ast }\mathcal{F}^{0}+%
\frac{1}{3}({c}yyy)g_{AB}\mathcal{F}^{A}\wedge {}^{\ast }\mathcal{F}^{B}%
\Bigg)  \notag \\
& +\frac{1}{6}\Big(3({c}x)_{AB}\mathcal{F}^{A}\wedge \mathcal{F%
}^{B}-3({c}xx)_{A}\mathcal{F}^{A}\wedge \mathcal{F}^{0}+({c}xxx)\mathcal{%
F}^{0}\wedge \mathcal{F}^{0}\Big)\Bigg]\;,  \label{Vector}
\end{align}%
where we are using the shorthand notation $({c}yyy)={c}%
_{ABC}y^{A}y^{B}y^{C}$, $({c}yy)_{A}={c}_{ABC}y^{B}y^{C} $ etc. 
When written in components the terms in the last line pick up an overall factor of $\varepsilon$, see appendix A.

The above action describes a theory of 4D, ${\cal N} = 2$ supergravity coupled to
vector-multiplets with Lorentzian spacetime signature if $\varepsilon =-1$
and Euclidean spacetime signature if $\varepsilon =+1$ \cite{mohaupt3}.
The scalar fields form a non-linear sigma model into a $2h_{1,1}$-dimensional 
projective special $\varepsilon$-K\"ahler manifold $M_{vector}$.
The four-dimensional spacetime metric is related to the ten- and even-dimensional spacetime metrics according to \eqref{metrics}.

In the rest of this section we explain how \eqref{Vector} is constructed term-by-term.

\subsection{Einstein-Hilbert term}

Consider the Einstein-Hilbert term and the kinetic term for
dilaton in the 10D action (\ref{10Dact}) 
\begin{equation}
S_{EH+\phi }^{10}=\int_{M_{10}} \left[ \frac{1}{2}{}^{\ast }R_{10}-\frac{9}{16}d\log
\phi \wedge {}^{\ast }d\log \phi \right] \;. \notag % 
\end{equation}%
Substituting the expression for the 10D Ricci scalar (\ref{exp}) into this action we find
\begin{align}
S_{EH+\phi }^{10}& =\int_{M_{10}} \Bigg[\frac{1}{2}{}^{\ast }R'_{4}-\frac{9}{16}d\log
\phi \wedge {}^{\ast }d\log \phi - dz^{\alpha }\wedge b_{\alpha
}\wedge {}^{\ast }(d\bar{z}^{\beta }\wedge \bar{b}_{\beta })  \notag \\
& \hspace{6em}+\frac{1}{2}dM^{A}\wedge V^{A}\wedge \left( {}^{\ast
}(dM^{B}\wedge {V}^{B})+\frac{1}{2}dM^{B}\wedge {V}^{B}\wedge J\right) \Bigg]%
\;, \notag
\end{align}%
where we have made use of the component expressions \eqref{eq:Comp1}, \eqref{eq:Comp2} and \eqref{hypermetric}.
We now integrate over the Calabi-Yau manifold to obtain the four-dimensional  action 
\begin{equation}
S_{EH+\phi }^{4}=\int_{M_{4}}\mathcal{V}\Bigg[\frac{1}{2}{}^{\ast }R'_{4}-%
\frac{9}{16}d\log \phi \wedge {}^{\ast }d\log \phi -\Galt_{\alpha
\bar{\beta} }dz^{\alpha }\wedge {}^{\ast }d\bar{z}^{\bar{\beta} }+\frac{1}{2}\left( {G_{AB}}+%
\frac{\mathcal{K}_{AB}}{\mathcal{V}}\right) dM^{A}\wedge {}^{\ast }dM^{B}%
\Bigg]\;. \notag
\end{equation}%
In order to write the
action in the Einstein frame we perform the Weyl rescaling 
\begin{equation}
(g'_4)_{\mu \nu }=\mathcal{V}^{-1}(g_4)_{\mu \nu }\;.  \label{Weyl}
\end{equation}%
Notice that in four-dimensions 
\begin{equation*}
\sqrt{g_4'}=\mathcal{V}^{-2}\sqrt{g_4}\;,
\qquad \sqrt{g'_4}g_4'{}^{\mu \nu }= \mathcal{V}^{-1}\sqrt{g_4}g_4^{\mu \nu }\;,
\qquad 
\sqrt{g_4'} g_4'^{\mu \nu }g_4^{\prime \rho \sigma }=\sqrt{g_4}g_4^{\mu \nu }g_4^{\rho \sigma }\;.
\end{equation*}%
After this transformation the 10D and 4D metrics are related via \eqref{metrics}.
The action is now written in the Einstein frame
\begin{align}
S_{EH+\phi }^{4}& =\int_{M_{4}}\Bigg[\frac{1}{2}{}^{\ast }R_{4}-\frac{3}{4}%
d\log \mathcal{V}\wedge {}^{\ast }d\log \mathcal{V}-\frac{9}{16}d\log \phi
\wedge {}^{\ast }d\log \phi   \notag \\
& \hspace{6em}-\Galt_{\alpha \bar{\beta} }dz^{\alpha }\wedge {}^{\ast }d%
\bar{z}^{\bar{\beta} }+\frac{1}{2}\left( {G_{AB}}+\frac{\mathcal{K}_{AB}}{\mathcal{%
V}}\right) dM^{A}\wedge {}^{\ast }dM^{B}\Bigg]\;. \notag
\end{align}%
Using the fact that $\mathcal{V}=\frac{1}{3!}\mathcal{K}$ and $d\mathcal{V}=%
\frac{1}{2}\mathcal{K}_{A}dM^{A}$ this can be written as
\begin{align}
S_{EH+\phi }^{4} &= \int_{M_{4}}\Bigg[\frac{1}{2}{}^{\ast }R_{4}-\frac{9}{16}%
d\log \phi \wedge {}^{\ast }d\log \phi -\Galt_{\alpha \bar{\beta}
}dz^{\alpha }\wedge {}^{\ast }d\bar{z}^{\bar{\beta} }  \notag \\
&\hspace{10em}-\frac{1}{2}\left( {G_{AB}}+\frac{9}{4}\frac{\mathcal{K}_{AB}}{\mathcal{K}%
^{2}}\right) dM^{A}\wedge {}^{\ast }dM^{B}\Bigg]\;. \notag
\end{align}%
Let us now make the field redefinition 
\begin{equation}
M^{A}=\sqrt{2}\phi ^{-3/4}v^{A}\;.  \label{Mv}
\end{equation}%
Since $\mathcal{K}$ is homogeneous of degree three in $M^A$ we have 
\[
	\mathcal{K}(M)=2\sqrt{2}\mathcal{K}(v)\phi ^{-9/4}\;,
	\qquad
	\mathcal{K}_{A}(M)=2 \mathcal{K}_{A}(v)\phi ^{-3/2}\;,
	\qquad 
	\text{etc}.  
\]
The action then takes the
form 
\begin{equation}
S_{EH+\phi }^{4}=\int_{M_{4}}\Bigg[\frac{1}{2}{}^{\ast }R_{4}-\frac{1}{2}{%
G_{AB}}(v)dv^{A}\wedge {}^{\ast }dv^{B}-\frac{1}{4}d\varphi \wedge {}^{\ast
}d\varphi -\Galt_{\alpha \bar{\beta} }(z,\bar{z})dz^{\alpha }\wedge {}^{\ast }d\bar{%
z}^{\beta }\Bigg]\;,  \label{gs}
\end{equation}%
where we have defined 
\begin{equation}
\varphi =\log \left( 2\mathcal{V}(v)\phi ^{-3}\right) \;.  \label{eq:Varphi}
\end{equation}%
Since there is are no factors of $\varepsilon$ in the action \eqref{gs} it is the same in both Lorentzian and Euclidean spacetime signatures.

The first term in \eqref{gs} is simply the four-dimensional Einstein-Hilbert term that appears in \eqref{Vector}. 
The second term contributes to the scalar sigma model appearing in \eqref{Vector}, which we will discuss next. The last two terms contribute to the action of the hyper-multiplets and will be dealt with in section \ref{sec:hyper}.

\subsection{Sigma model}

We now consider the contribution from the $H^2$ cohomology sector of the $B_2$ field in the 10D action  (\ref{10Dact})
\begin{equation*}
S_{H^{2}(B_{2})}^{10}=\int \left[ \frac{1}{2}\varepsilon \,\phi ^{-\frac{3}{2%
}}H_{3}\wedge {}^{\ast }H_{3}\Big|_{H^{2}}\right] \;.
\end{equation*}%
We anticipate that the overall factor of $\varepsilon$ will remain in place after dimensional reduction over the Calabi-Yau manifold.

Substituting $H_{3}\big|%
_{H^{2}}=da^{A}\wedge V^{A}$ into the above action and integrating over the Calabi-Yau three-fold results in the 4D effective action
\begin{equation}
S_{H^{2}(B_{2})}^{4}
=\int_{M_{4}}\left[ \varepsilon \,\phi ^{-\frac{3}{2}}\mathcal{V}%
G_{AB}(M)da^{A}\wedge {}^{\ast }da^{B} \right]\;. \notag
\end{equation}%
Performing the Weyl rescaling (\ref{Weyl}) and making the field
redefinition (\ref{Mv}) we obtain 
\begin{equation}
S_{H^{2}(B_{2})}^{10}=\int_{M_{4}}\left[\varepsilon \,\frac{1}{2}%
G_{AB}(v)da^{A}\wedge {}^{\ast }da^{B} \right]\;. \label{eq:gaa}
\end{equation}%
As expected, we the overall factor of $\varepsilon$ has survived to the 4D action.

One may combine \eqref{eq:gaa} with the $H^2$ contribution from (\ref{gs}) (which is given by the second term) to obtain the enlarged sigma model 
\begin{equation}
S_{H^{2}(EH+\phi)}^{4} + S_{H^{2}(B_{2})}^{4}=\int_{M_{4}}\left[ -\frac{1}{2}%
G_{AB}(v)\left( dv^{A}\wedge {}^{\ast }dv^{B}-\varepsilon \,da^{A}\wedge
{}^{\ast }da^{B}\right) \right] \;.  \notag
\end{equation}
In order to compare this expression with the existing literature it is convenient
to make the field redefinition 
\begin{equation}
v^{A}=\frac{1}{2^\frac16} y^{A}\;,\qquad  a^{A}=-\frac{1}{2^\frac16}x^{A}\;,\qquad \mathcal{K}_{ABC}=c_{ABC}\;.
\label{Transformation0}
\end{equation}%
Due to the homogeneity properties of $G_{AB}$ the factors of $2^\frac16$ are irrelevant  to the above action, however they will be useful later when considering terms involving the gauge fields. After the field redefinitions \eqref{Transformation0} the action is given by 
\begin{equation}
S_{H^{2}(EH+\phi)}^{4} + S_{H^{2}(B_{2})}^{4}=\int_{M_{4}}\Big[-g_{AB}(y)\left(
dx^{A}\wedge {}^{\ast }dx^{B}-\varepsilon \,dy^{A}\wedge {}^{\ast
}dy^{B}\right) \Big]\;, \label{SigmaModel}
\end{equation}%
where we have defined
\begin{equation}
g_{AB}(y):=-\varepsilon \,\frac{1}{2}G_{AB}(y)=\varepsilon \frac{3}{2}\left( 
\frac{({c}y)_{AB}}{({c}yyy)}-\frac{3}{2}\frac{({c}y)_{A}({c}y)_{B}}{({c}%
yyy)^{2}}\right) \;. \label{gAB}
\end{equation}%
This agrees with the expression for the vector-multiplet sigma model for
Lorentzian or Euclidean spacetime signatures given in \cite{mohaupt3}. It
corresponds to the second term in the action (\ref{Vector}).

We remark that the target manifold ${M}_{vector}$ described by the sigma model \eqref{SigmaModel} is a $2h_{1,1}$-dimensional projective special $\varepsilon $-K\"{a}hler manifold. 
In order to expose this property one may define the $\varepsilon $-complex coordinates%
\footnote{
In \cite{inst1} different conventions are used for the $\varepsilon$-complex coordinates. They can be matched with the conventions used here by setting $y^A \to -y^A$, $c_{ABC} \to -c_{ABC}$ and $\epsilon_{\mu\nu\rho \sigma} \to -\epsilon_{\mu\nu\rho \sigma}$.} 
$w^{A}=x^{A}+i_{\varepsilon }y^{A}$, where $i_{\varepsilon }$ is the $\varepsilon $-complex
unit. In these coordinates the $\varepsilon$-K\"{a}hler potential is given by
\begin{equation}
K=-\log \mathcal{V}(y)\;,\qquad \mathcal{V}(y)=\frac{1}{3!}{c}%
_{ABC}y^{A}y^{B}y^{C}\;, \notag
\end{equation}%
where it is understood that $y^A = \text{Im}(w^A)$.
The $\varepsilon $-holomorphic
prepotential on the corresponding conic-affine special $\varepsilon $-K\"{a}%
hler manifold is given by $F=-\frac{1}{6}{c_{ABC}Z^{A}Z^{B}Z^{C}}/{Z^{0}}$,
where $(Z^{0},\ldots ,Z^{h_{1,1}})$ are homogeneous special $\varepsilon $%
-holomorphic coordinates satisfying $w^A = Z^A / Z^0$. Since the coefficients $c_{ABC}$ are real there is a 1-1 correspondence between holomorphic prepotentials $F_{\varepsilon = -1}$ and para-holomorphic prepotentials $F_{\varepsilon = +1}$, at least in the the context of dimensional reduction over a Calabi-Yau three-fold considered in this paper. This is not true for holomorphic and para-holomorphic functions in general.

\subsection{Gauge fields}

We now turn our attention to the terms involving gauge fields in \eqref{Vector}.
The starting point is the contribution from the $H^{2}$ cohomology sector of the $A_{3}$ 
fields to the non-topological part of the 10D action 
\begin{equation}
S_{H^{2}(A_{3})}^{10}=\int_{M_{10}}\left[ -\frac{1}{2}\phi ^{\frac{3}{4%
}}(F_{4}+dV\wedge B_{2})\wedge {}^{\ast }(F_{4}+dV\wedge B_{2})\Big|_{H^{2}}%
\right] \;. \notag
\end{equation}%
The individual terms decompose according to
\begin{equation}
	F_4 \big|_{H^{2}} = {\cal F}^A \wedge V^A \;,
	\qquad
	B_2 \big|_{H^{2}} = a^A V^A \;,
	\label{eq:H2decomp}
\end{equation}
where we have defined ${\cal F}^A = d{\cal A}^A$.
Plugging this into the action we get 
\begin{align}
S_{H^{2}(A_{3})}^{10}& =\int_{M_{4}}\left[ -\frac{1}{2}\phi ^{\frac{3%
}{4}}(\mathcal{F}^{A}+a^{A}dV)\wedge {}^{\ast }\left[ (\mathcal{F%
}^{B}+a^{B}dV)\right] \right] \int_\CY V^A \wedge {}^* V^B \notag  \;.
\end{align}
Integrating over the Calabi-Yau three-fold and making the field redefinition (\ref{Mv}) gives us
\begin{align}
S_{H^{2}(A_{3})}^{4}
& =\int_{M_{4}}\left[ -\frac{\sqrt{2}}{3!}\mathcal{K}(v)G_{AB}(v)(\mathcal{F}%
^{A}+a^{A}dV)\wedge {}^{\ast }(\mathcal{F}^{B}+a^{B}dV)\right] \;. \label{eq:H2A}
\end{align}
Notice that this action is invariant under Weyl transformations. 

Consider now the contribution from the 11D Kaluza-Klein vector kinetic term
\begin{equation}
S_{V}^{10}=\int_{M_{10}} \left[\frac{1}{4}\varepsilon \,\phi ^{\frac{9}{4}}dV\wedge
{}^{\ast }dV\right] \;, \notag
\end{equation}%
which upon integrating over the Calabi-Yau manifold and making the field redefinition (\ref{Mv}) gives
\begin{equation}
S_{V}^{4} 
=\int_{M_{4}}\left[ \varepsilon \,\frac{\sqrt{2}}{2}\frac{1}{3!}\mathcal{K}(v)\mathcal{F}%
^{0}\wedge {}^{\ast }\mathcal{F}^{0} \right]\;, \notag
\end{equation}%
where we have defined $\mathcal{F}^{0}=dV$. Combining this with \eqref{eq:H2A} we get 
\begin{align}
\hspace{-2em}S_{H^{2}(A_{3})}^{4}+S_{V}^{4}  
& =\int_{M_{4}}\Bigg[\sqrt{2}\left( \varepsilon \,\frac{1}{12}(\mathcal{K}%
vvv)-\frac{1}{6}(\mathcal{K}vvv)G_{AB}a^{A}a^{B}\right) \mathcal{F}%
^{0}\wedge {}^{\ast }\mathcal{F}^{0}  \notag \\
& \hspace{4em}-\frac{\sqrt{2}}{3}(\mathcal{K}vvv)G_{AB}a^{B}\mathcal{F}%
^{A}\wedge {}^{\ast }\mathcal{F}^{0}-\frac{\sqrt{2}}{6}(\mathcal{K}vvv)G_{AB}%
\mathcal{F}^{A}\wedge {}^{\ast }\mathcal{F}^{B}\Bigg]\;, \notag
\end{align}%
where $({\cal K}vvv):={\cal K}_{ABC} v^A v^B v^C = {\cal K}(v)$.
We now make the field redefinitions (\ref{Transformation0}) and \eqref{gAB} along with 
\begin{equation}
	\mathcal{F}^{A}=\frac{1}{2^{\frac{1}{6}}}\mathcal{F}^{\prime A} \;,
	\label{eq:Frescale}
\end{equation}
to get (dropping the primes) 
\begin{align}
& S_{H^{2}(A_{3})}^{4}+S_{V}^{4}  \notag \\
& =\int_{M_{4}}\varepsilon \,\Bigg[\frac{1}{2}({c}yyy)\left( \frac{1}{6}+%
\frac{2}{3}(gxx)\right) \mathcal{F}^{0}\wedge {}^{\ast }\mathcal{F}^{0}-%
\frac{2}{3}({c}yyy)(gx)_{A}\mathcal{F}^{A}\wedge {}^{\ast }\mathcal{F}^{0}+%
\frac{1}{3}({c}yyy)g_{AB}\mathcal{F}^{A}\wedge {}^{\ast }\mathcal{F}^{B}%
\Bigg]\;, \notag
\end{align}%
which corresponds to the second line of \eqref{Vector}.

Finally, we consider the contributions of the $H^2$-sector to the topological term in the 10D action
\begin{equation}
S_{H^{2}(top)}^{10}=\int_{M_{10}}\left[ -\frac{\sqrt{2}}{2}\left( {F}%
_{4}+dV\wedge B_{2}\right) \wedge {F}_{4}\wedge B_{2}-\frac{\sqrt{2}}{6}%
dV\wedge B_{2}\wedge dV\wedge B_{2}\wedge B_{2} \Big|_{H^{2}}\right]\;. \notag
\end{equation}%
Substituting in \eqref{eq:H2decomp} we find
\begin{align}
S_{H^{2}(top)}^{10} & =\int_{M_{4}}\left[ -\frac{\sqrt{2}}{2}\left( \mathcal{F}%
^{A}+\mathcal{F}^{0}a^{A}\right) \wedge \mathcal{F}^{B}a^{C}-\frac{\sqrt{2}}{%
6}\mathcal{F}^{0}a^{A}\wedge \mathcal{F}^{0}a^{B}a^{C}\right]
\int_{\CY} V^{A}\wedge V^{B}\wedge V^{C}  \;, \notag
\end{align}%
which upon integration gives us
\begin{align}
S_{H^{2}(top)}^{4} 
& =\int_{M_{4}}\left[ -\frac{\sqrt{2}}{2}\left( (\mathcal{K}a)_{AB}\mathcal{F%
}^{A}\wedge \mathcal{F}^{B}+(\mathcal{K}aa)_{A}\mathcal{F}^{A}\wedge 
\mathcal{F}^{0}+\frac{1}{3}(\mathcal{K}aaa)\mathcal{F}^{0}\wedge \mathcal{F}%
^{0}\right) \right] \;. \notag
\end{align}%
After making the field redefinitions (\ref{Transformation0}) and \eqref{eq:Frescale} we get
\begin{equation*}
S_{H^{2}(top)}^4=\int_{M_{4}}\left[ \frac{1}{6}\left( 3({c}x)_{AB}\mathcal{F}%
^{A}\wedge \mathcal{F}^{B}-3({c}xx)_{A}\mathcal{F}^{A}\wedge \mathcal{F}%
^{0}+({c}xxx)\mathcal{F}^{0}\wedge \mathcal{F}^{0}\right) \right] \;,
\end{equation*}%
which produces the last line of \eqref{Vector}.

\section{4D hyper-multiplets}
\label{sec:hyper}

In this section we will consider the hyper-multiplet part of the reduction of the 10D action (\ref%
{10Dact}) over a Calabi-Yau three-fold. We will show that the contribution from the 
$H^{3}$-sector, the dilaton and $H^{0}$-sector of the $B_{2}$ field
results in the 4D effective action with spacetime signature $(\varepsilon,+,+,+)$
\begin{align}
& S_{hyper}^{4}=\int_{M_{4}}\Bigg[-\Galt_{\alpha \bar{\beta}}dz^{\alpha }\wedge
{}^{\ast }d\bar{z}^{\bar{\beta} }-\frac{1}{4}d\varphi \wedge {}^{\ast }d\varphi 
\notag \\
& \hspace{4em}-\;e^{-2\varphi }\left( d\tilde{\phi}+\frac{1}{2}\left( \zeta
^{I}d\tilde{\zeta}_{I}-\tilde{\zeta}_{I}d\zeta ^{I}\right) \right) \wedge
{}^{\ast }\left( d\tilde{\phi}+\frac{1}{2}\left( \zeta ^{I}d\tilde{\zeta}%
_{I}-\tilde{\zeta}_{I}d\zeta ^{I}\right) \right)  \notag \\
& \hspace{4em}-\;\lambda \frac{1}{2}e^{-\varphi }\left( \mathcal{I}%
_{IJ}d\zeta ^{I}\wedge {}^{\ast }d\zeta ^{J}+\mathcal{I}^{IJ}\left( d\tilde{%
\zeta}_{I}+\mathcal{R}_{IK}d\zeta ^{K}\right) \wedge {}^{\ast }\left( d%
\tilde{\zeta}_{I}+\mathcal{R}_{IK}d\zeta ^{K}\right) \right) \Bigg]\;,
\label{Hyper}
\end{align}%
where $\lambda = -1$ and the coupling matrices $\Galt_{\alpha \bar{\beta}}, {\cal I}_{IJ}, {\cal R}_{IJ}$ depend on $z^\alpha$. Note that in our conventions ${\cal I}_{IJ}$ is negative definite.

For the case $\varepsilon = -1$ the above action, when combined with the vector-multiplet action \eqref{Vector}, describes the bosonic part of ${\cal N} = 2$ hyper-multiplets coupled to supergravity with Lorentzian spacetime signature \cite{Ferrara:1989ik}. For the case  $\varepsilon = +1$ it is expected to describe the bosonic part of ${\cal N} = 2$ hyper-multiplets coupled to supergravity with Euclidean spacetime signature. Indeed, it is clear that this action can be obtained by the reduction of 5D, ${\cal N} = 2$ local hyper-multiplets over a timelike circle, since the bosonic part of the hyper-multiplet action does not change upon dimensional reduction.

The action \eqref{Hyper} describes a non-linear sigma model into a positive-definite $(4h_{2,1} + 4)$-dimensional quaternionic K\"ahler target manifold ${M}_{hyper}$ \cite{Ferrara:1989ik}. 
Notice that the parameter $\varepsilon$ does not
appear in front of any terms in the action, nor does it appear in the
definitions of the scalar fields or coupling matrices. Therefore the coupling matrices, and, hence, the scalar target geometry (quaternionic K\"ahler) is the same regardless of whether the reduction from 11D to 10D was performed over a spacelike or timelike circle.

We remark that the bosonic sector of 3D, ${\cal N} = 2$ local Euclidean hyper-multiplets obtained by the dimensional reduction (followed by dualisation) of 4D, ${\cal N} = 2$ vector-multiplets over time takes the same form as the above action with $\lambda = +1$. In this case the target manifold has split signature and the metric is para-quaternionic K\"ahler \cite{Vaughan:2012, Cortes:2015}. We also anticipate that the above action with $\lambda = +1$ can be obtained by the reduction of 10D type IIA${}^*$ supergravity with Lorentzian spacetime signature, as described in \cite{Hull:1998vg}, over a Calabi-Yau three-fold. This is because the sign flip in front of the $G_4^2$ term in the IIA${}^*$ action, which is given by expression (4.8) of \cite{Hull:1998vg}, corresponds to setting $\lambda = +1$ in \eqref{Hyper}. On the other hand, the sign flips in front of the $H^2$ and the topological terms of the IIA${}^*$ action will be compensated by a sign flip in the Hodge dualisation procedure when $\varepsilon = -1$.

The first line in (\ref{Hyper}) is taken from the $H^{3}$ and dilaton terms
in the gravity sector (\ref{gs}). In the remainder of this section we will
explain the origins of the second and third lines. We will use the conventions 
for special K\"ahler geometry given in appendix A.

\subsection{$\protect\zeta^I,\tilde{\protect\zeta}_I$ terms}

Let us consider the non-topological part of the $H^{3}$-sector of the 10D
action 
\begin{equation}
S_{H^{3}(ntop)}^{10}=\int_{M_{10}}\left[ -\frac{1}{2}\phi ^{\frac{3}{4}}{F}%
_{4}\wedge {}^{\ast }{F}_{4}\Big|_{H^{3}}\right] \;. \notag
\end{equation}%
Substituting in the expression (\ref{F4decomp}) for $F_{4}\big|_{H^{3}}=d%
\check{A}$ into the action gives
\begin{align*}
S_{H^{3}(ntop)}^{10}& =\int_{M_{4}}\left[ -\phi ^{\frac{3}{4}}
P^{I}\wedge {}^{\ast }\bar{P}^{J} \right]
\int_\CY \Phi_I \wedge {}^*\bar{\Phi}_J
+ \int_{M_{4}}\left[
\phi ^{\frac{3}{4}} \bar{Q}\wedge {}^{\ast }Q \right]  
\int_\CY \bar{\Omega} \wedge {}^* {\Omega} \;.
\end{align*}
Integrating over the Calabi-Yau three-fold we obtain the 4D action 
\begin{align}
S_{H^{3}(ntop)}^{4}& =\int_{M_{4}}\left[ \phi ^{\frac{3}{%
4}}\left( \mathcal{M}_{I\bar{J}}P^{I}\wedge {}^{\ast }\bar{P}^{J}-(\bar{X}NX)\bar{Q%
}\wedge {}^{\ast }Q\right) \right]  \notag \\
& =\int_{M_{4}}\Bigg[  \frac{\sqrt{2}}{2}\phi ^{\frac{3}{4}}%
\mathcal{I}^{IJ}\left(d\tilde{\zeta}_{I} + \mathcal{N}_{IK}d\zeta ^{K}  \right) \wedge {}^{\ast }\left( d\tilde{\zeta}_{J} + \bar{\mathcal{N}}_{JL}d\zeta ^{L}  \right)\Bigg] , \notag
\end{align}%
where in the last line we used the expression for $\mathcal{I}^{-1}$ given
in \eqref{eq:SGconventions}. We now make the Weyl transformation (%
\ref{Weyl}) to get 
\begin{equation}
S_{H^{3}(ntop)}^{4}=\int_{M_{4}}\Bigg[  \frac{1}{2}e^{-\varphi }%
\mathcal{I}^{IJ}\left( d\tilde{\zeta}_{I} + \mathcal{N}_{IK}d\zeta ^{K} \right) \wedge {}^{\ast }\left( d\tilde{\zeta}_{J} + \bar{\mathcal{N}}_{JL}d\zeta ^{L} \right) \Bigg] \;. \notag
\end{equation}%
where $\varphi $ was defined in (\ref{eq:Varphi}). Substituting ${\cal N}_{IJ} = {\cal R}_{IJ} + i{\cal I}_{IJ}$ followed by a straight-forward rewriting gives the third line of (\ref{Hyper}).

\subsection{$\tilde{\protect\phi}$ term}
\label{sec:axion}

We now consider the topological part of the $H^{3}$-sector of the 10D
action 
\begin{equation}
S_{H^{3}(top)}^{10}=\int_{M_{10}}\Bigg[-\frac{\sqrt{2}}{2}F_{4}\wedge
F_{4}\wedge B_{2}\Big|_{H^{3}}\Bigg]\;. \notag
\end{equation}%
Substituting in (\ref{F4decomp}) we find
\begin{align}
S_{H^{3}(top)}^{10}& =\int_{M_{4}}\left[ -\sqrt{2}\, \mathcal{B}_{2}\wedge P^{I}\wedge \bar{P}^{J}\right] \int_\CY \Phi_I \wedge \bar{\Phi}_J + \int_{M_{4}}\left[\sqrt{2}\,\mathcal{B}_{2}\wedge\bar{Q} \wedge Q \right] \int_\CY \bar{\Omega} \wedge \Omega \notag \;. 
\end{align}%
Making use of \eqref{eq:degree} we integrate over the Calabi-Yau three-fold to obtain 
\begin{align}
S_{H^{3}(top)}^{4}& =\int_{M_{4}}\left[ i \sqrt{2}\mathcal{B}%
_{2}\wedge \left( \mathcal{M}_{I\bar{J}}P^{I}\wedge \bar{P}^{J}-(\bar{X}NX)\bar{Q}%
\wedge Q\right) \right]  \notag \\
& =\int_{M_{4}}\Big[i  \mathcal{B}_{2}\wedge \mathcal{I}^{IJ}\left(
d\tilde{\zeta}_{I} + \mathcal{N}_{IK}d\zeta ^{K} \right)
\wedge \left( d\tilde{\zeta}_{J} + \bar{\mathcal{N}}_{JL}d\zeta ^{L} \right)\Big]  \notag \\
& =\int_{M_{4}}\Big[-2 \mathcal{B}_{2}\wedge d\zeta ^{I}\wedge d\tilde{%
\zeta}_{I}\Big]\;, \notag
\end{align}%
where in the last line we used ${\cal N}_{IJ} = {\cal R}_{IJ} + i{\cal I}_{IJ}$. Note that this term is invariant under Weyl
rescalings. Integrating by parts and adding the contribution from the $H^{0}$-sector
of the $B_{2}$ field gives 
\begin{equation}
S_{H^{3}(top)}^4 + S^4_{H^{0}(B_{2})} =\int_{M_{4}}\Bigg[2 \mathcal{H}%
_{3} \wedge \zeta ^{I} d\tilde{\zeta}_{I}+\varepsilon 
e^{2\varphi }\mathcal{H}_{3}\wedge {}^{\ast }\mathcal{H}_{3}\Bigg]\;. \notag
\end{equation}
We now dualise the three-form $\mathcal{H}_{3}$ by adding the Lagrange
multiplier 
\begin{equation}
S_{Lm}^{4}=\int_{M_{4}} \Bigg[ 2 \mathcal{H}_{3}\wedge d\left( 
\tilde{\phi} - \frac{1}{2}\zeta ^{I}\tilde{\zeta}_{I}\right) \Bigg] \;. \notag
\end{equation}%
Solving the Euler-Lagrangian equations of $%
S_{H^{3}(top)}^4 + S^4_{H^{0}(B_{2})}+S_{Lm}^{4}$ for $\mathcal{H}_{3}$ gives 
\begin{equation}
{}^* \mathcal{H}_{3}= -\varepsilon e^{-2\varphi }
\left( d\tilde{\phi}+\frac{1}{2}(\zeta ^{I}d\tilde{\zeta}_{I}-\tilde{\zeta}%
_{I}d\zeta ^{I})\right) \;. \notag
\end{equation}%
Substituting back into the action we get 
\begin{equation*}
S_{H^{3}(top)}^4 + S^4_{H^{0}(B_{2})} + S_{Lm}^{4}=\int_{M_{4}}\left[ -e^{-2\varphi
}\left( d\tilde{\phi}+\frac{1}{2}(\zeta ^{I}d\tilde{\zeta}_{I}-\tilde{\zeta}%
_{I}d\zeta ^{I})\right) \wedge {}^{\ast }\left( d\tilde{\phi}+\frac{1}{2}%
(\zeta ^{I}d\tilde{\zeta}_{I}-\tilde{\zeta}_{I}d\zeta ^{I})\right) \right],
\end{equation*}%
where we have used the fact that $* * \alpha = - \varepsilon \alpha$ for any one-form or three-form on $M_4$. This produces the second line of (\ref{Hyper}).

\subsection*{Acknowledgements}

We would like to thank Thomas Mohaupt and Chris Hull for useful discussions. We would also like to thank the referees for useful comments and suggestions. The work of W.S.\ is supported in part by the National Science Foundation under grant number PHY- 1415659. The work of O.V.\ is supported by the German Science Foundation (DFG) under the Collaborative Research Center (SFB) 676 ``Particles, Strings and the Early Universe.''

\appendix

\section{Conventions and identities}

\label{sec:Conventions}

Consider an $m$-dimensional pseudo-Riemannian manifold with signature $(k,\ell),$ where $k$ represents the number of timelike dimensions. We take the epsilon symbol and tensor respectively to be 
\[
	\epsilon_{12\ldots m} = 1\;, 
	\qquad
	\varepsilon_{12\ldots m} = \sqrt{|g|} \epsilon_{12\ldots m}  \;.
\]
Note that the epsilon tensor $\varepsilon_{\mu_1\ldots\mu_m}$ will always be written with indices to avoid confusion with the parameter $\varepsilon = \pm 1$ introduced in \eqref{eq:Epsilon}.
One may use the metric to raise the indices of the epsilon tensor
\[
	\varepsilon^{\mu_1 \ldots \mu_m} := g^{\mu_1 \nu_1}\ldots g^{\mu_m \nu_m} \varepsilon_{\nu_1 \ldots \nu_m} = (-1)^k \sqrt{|g|}^{-1} \epsilon_{\mu_1 \ldots \mu_m}.
\] 
It follows that 
\[
	dx^{\mu_1}\wedge \ldots \wedge dx^{\mu_m} = (-1)^k \sqrt{|g|} \varepsilon^{\mu_1 \ldots \mu_m} dx^1 \wedge \ldots \wedge dx^m = (-1)^k \sqrt{|g|} \varepsilon^{\mu_1 \ldots \mu_m} d^mx \;,
\]
and
\[
	\varepsilon_{\mu_1\ldots \mu_p \rho_{p+1} \ldots \rho_{m}}
	\varepsilon^{\nu_1\ldots \nu_p \rho_{p+1} \ldots \rho_{m}}
	= (-1)^k p! (m-p)! \delta^{[\nu_1}_{[\mu_1}\ldots\delta^{\nu_p]}_{\mu_p]}\;.
\]
Differential $p$-forms are expanded according to
\[
	\alpha_p = \frac{1}{p!} (\alpha_p)_{\mu_1 \ldots \mu_p} dx^{\mu_1} \wedge \ldots \wedge dx^{\mu_p} \;.
\]
The Hodge star is defined by
\[
	{}^* \alpha_p = \frac{1}{p!(m - p)!} (\alpha_p)_{\mu_1 \ldots \mu_p} \varepsilon^{\mu_1 \ldots \mu_p}_{\phantom{\mu_1 \ldots \mu_p} \nu_{p+1} \ldots \nu_{m}} dx^{\nu_{p+1}} \wedge \ldots \wedge dx^{\nu_m} \;,
\]
and therefore
\[
	{\alpha}_p \wedge {}^* \beta_p = \frac{1}{p!} (\alpha_p)_{\mu_1 \ldots \mu_p} (\beta_p)^{\mu_1 \ldots \mu_p} \sqrt{|g|} d^mx \;.
\]
Notice that 
\[
	\alpha_p \wedge \gamma_{(m-p)} = \frac{1}{p!(m - p!)}(\alpha_p)_{\mu_1 \ldots \mu_p} (\gamma_{(m-p)})_{\nu_{p+1}\ldots \nu_m} \varepsilon^{\mu_1 \ldots \mu_p \nu_{p+1}\ldots \nu_m} (-1)^k \sqrt{|g|} d^mx \;,
\]
so, for example, the last line of \eqref{10Dact} is written in components as
\begin{multline}
	\int_{M_{10}} d^{10}w \sqrt{|g_{10}|} \varepsilon^{\hat{\mu}_1 \ldots \hat{\mu}_{10}} \Bigg[ - \varepsilon \, \frac{\sqrt{2}}{48^2} \left(F_{\hat{\mu}_1 \hat{\mu}_2 \hat{\mu}_3 \hat{\mu}_4} + 6V_{\hat{\mu}_1 \hat{\mu}_2} B_{\hat{\mu}_3 \hat{\mu}_4}\right) F_{\hat{\mu}_5\hat{\mu}_6\hat{\mu}_6\hat{\mu}_7}B_{\hat{\mu}_9\hat{\mu}_{10}}  \\
	-\varepsilon \, \frac{\sqrt{2}}{192} V_{\hat{\mu}_1\hat{\mu}_2} B_{\hat{\mu}_3\hat{\mu}_4} V_{\hat{\mu}_5\hat{\mu}_6} B_{\hat{\mu}_7\hat{\mu}_8} B_{\hat{\mu}_9\hat{\mu}_{10}} \Bigg]\;. \label{eq:10Dlast}
\end{multline}

We use the following conventions for special K\"ahler geometry: 
\begin{align}
N_{IJ}& =2\text{Im}(F_{IJ})=\frac{1}{i}(F_{IJ}-\bar{F}_{IJ})  \notag \\
\mathcal{N}_{IJ}& =\bar{F}_{IJ}+i\frac{(NX)_{I}(NX)_{J}}{XNX}  \notag \\
\mathcal{I}_{IJ}& =\text{Im}(\mathcal{N}_{IJ})=-\frac{1}{2}N_{IJ}+\frac{1}{2}%
\frac{(NX)_{I}(NX)_{J}}{XNX}+\frac{1}{2}\frac{(N\bar{X})_{I}(N\bar{X})_{J}}{%
\bar{X}N\bar{X}}  \notag \\
\mathcal{R}_{IJ}& = \text{Re}(\mathcal{N}_{IJ}) = \frac12(F_{IJ} + \bar{F}_{IJ}) + \frac{i}2 \frac{(NX)_{I}(NX)_{J}}{XNX} - \frac{i}{2} \frac{(NX)_{I}(NX)_{J}}{XNX}  \notag \\ 
K_{I}& =\frac{\partial }{\partial X^{I}}\log (\bar{X}NX)=\frac{(N\bar{X})_{I}%
}{\bar{X}NX}  \notag \\
\mathcal{I}^{IJ}& =-2N^{IJ}+2\frac{X^{I}\bar{X}^{J}}{\bar{X}NX}+2\frac{\bar{X%
}^{I}{X}^{J}}{\bar{X}NX}=2N^{IK}\left(-\delta^J_K+\bar{K}_K\bar{X}^J+{K}_K{X}^J\right)  \notag \\
\mathcal{M}_{I\bar{J}}& =-N_{IJ}+\frac{(N\bar{X})_{I}(N{X})_{J}}{\bar{X}NX} 
\notag \\
\mathcal{F}_{IJ}& =F_{IJ}\;, \label{eq:SGconventions}
\end{align}%
where $(NX)_I = N_{IJ} X^J$ and $\bar{X}NX = N_{IJ} \bar{X}^I X^J$ etc. 
The matrix $N_{IJ}$ has complex Lorentz signature and ${\cal I}_{IJ}$ is negative definite.
We will often omit writing indices explicitly when the meaning is clear from the order.
Some useful identities are 
\begin{equation}
\frac{\partial \bar{K}}{\partial \bar{X}}d\bar{X}=i(d\bar{\mathcal{F}})N^{-1}%
\bar{K}-\bar{K}(\bar{K}d\bar{X})  \label{Id1}
\end{equation}%
\begin{equation}
d\mathcal{I}^{-1}=2iN^{-1}(d\bar{\mathcal{F}})N^{-1}(Id-KX)+2N^{-1}\left( 
\frac{\partial \bar{K}}{\partial X}dX\right) \bar{X}+2(N^{-1}\bar{K})d\bar{X}%
(Id-\bar{K}\bar{X})+h.c. \label{Id2}
\end{equation}%
\begin{equation}
\mathcal{I}^{-1}(d\mathcal{N})\mathcal{I}^{-1}=\frac{4i}{\bar{X}NX}\Big[-(dX%
\bar{X}+\bar{X}dX)+(KdX)(X\bar{X}+\bar{X}X)\Big]+4N^{-1}\left( d\bar{%
\mathcal{F}}\right) N^{-1}\;.  \label{Id3}
\end{equation}

\section{Alternative calculation of $d\check{A}$}

In a previous version of this paper a different calculation for the exterior derivation of $\check{A}$ was presented, which closely followed the original calculation of \cite{Bodner}. The calculation that now appears in the main text is far more concise and does not involve evaluating differentials. We will include here our original calculation for the purpose of continuity with previous versions of this paper, and because it provides a complementary approach to this calculation using  complex forms.

We start by writing $\check{A}$ as 
\begin{equation}
\check{A}=\Psi (a\alpha +b\beta )+\bar{\Psi}(\bar{a}\alpha +\bar{b}\beta )=i%
\text{Im}(\Psi )2a\alpha +i\text{Im}(\Psi )(b-\bar{b})\beta +\text{Re}{\Psi }%
(b+\bar{b})\beta \;. \notag
\end{equation}%
Here $\Psi_I(x)$ are a complex fields and $(a^{IJ}(x)),(b_{J}^{I}(x))$ are complex matrices, where we have chosen $a$ to
be purely imaginary. Substituting in the expression for $\alpha _{I},\beta 
^{I}$ given in (\ref{eq:alpha}), (\ref{eq:beta}) we get 
\begin{equation}
\check{A}=i\Psi \Big((a\bar{\mathcal{F}}+b)N^{-1}\Omega -(a\mathcal{F}%
+b)N^{-1}\bar{\Omega}\Big)+h.c.\ \;. \notag
\end{equation}%
We now make the ansatz 
\begin{equation}
(a\mathcal{F}+b)N^{-1}\bar{\Phi}=0\;,\qquad (a\mathcal{F}+b)N^{-1}\bar{K}%
\bar{\Omega}\propto \bar{K}\bar{\Omega}\;, \notag
\end{equation}%
which can satisfied by setting 
\begin{equation}
(a\mathcal{F}+b)N^{-1}=d\bar{K}\bar{X}\qquad \Rightarrow \qquad b=d\bar{K}%
\bar{X}N-a\mathcal{F}\;. \notag
\end{equation}%
Substituting this into the expression for $\check{A}$ we get 
\begin{equation}
\check{A}=i\Psi \Big((-ia+d\bar{K}\bar{X})(\Phi +K\Omega )-d\bar{K}\bar{%
\Omega}\Big)+h.c.\ \;. \notag
\end{equation}%
It is convenient to choose $d=2^{\frac{1}{4}} N^{-1}$, in which case 
\begin{equation}
\check{A}=2^{\frac{1}{4}} i \Psi N^{-1}\left( \Phi -\bar{K}\bar{\Omega}\right) +h.c.\;.  \notag
\end{equation}%
Comparing with \eqref{eq:_Acheck} gives
\begin{align}
\zeta ^{I}& =\left( \text{Im}(\Psi )\mathcal{I}^{-1}\right)\hspace{-3pt}{}^I \;, 
\notag \\
\tilde{\zeta}_{I}& =\left( \text{Im}(\Psi )N^{-1}%
\big[(Id-2KX)\mathcal{F}+(Id-2\bar{K}\bar{X})\bar{\mathcal{F}}\big]+\text{Re}%
(\Psi )\right)\hspace{-3pt}{}_I \;. \label{eq:ZetaPsi}
\end{align}%
The complex fields $\Psi _{I}$ are related to the real fields $\zeta ^{I},%
\tilde{\zeta}_{I}$ by 
\begin{equation}
\Psi _{I}=\tilde{\zeta}_{I}+\mathcal{N}_{IJ}\zeta ^{J}\;.  \label{PsiZeta}
\end{equation}
Taking derivatives we find
\begin{equation}
\left( d{\Psi }+\tfrac{i}{2}(\Psi -\bar{\Psi})\mathcal{I}^{-1}d\mathcal{N}%
\right)\hspace{-3pt}{}_{I}=d\tilde{\zeta}_{I}+\mathcal{N}_{IJ}d\zeta ^{J}\;.
\label{eq:dPsi}
\end{equation}

We now take derivative of $\check{A}$ using the expressions \eqref{eq:_Acheck} and \eqref{eq:ZetaPsi}. 
After some simplifications using identities (\ref{Id1}) and (\ref{Id2}) along with \eqref{eq:Omega}  and \eqref{Phidecomp} we get 
\begin{align}
d\check{A}& =2^{\frac{1}{4}} i\left( d\Psi N^{-1}-(\Psi -\bar{\Psi}%
)N^{-1}\left( (KdX)+i(d\bar{\mathcal{F}})N^{-1}\right) \right)\hspace{-3pt}{}^I \Phi_I  \notag
\\
& \hspace{4em}-2^{\frac{1}{4}} i\left( d\Psi N^{-1}\bar{K}+(\Psi -%
\bar{\Psi})N^{-1}\left( \frac{\partial \bar{K}}{\partial X}dX\right) \right) 
\bar{\Omega}+h.c.\;. \notag
\end{align}%
Simplifying further using \eqref{Id3} we can write this more concisely as 
\begin{align}
d\check{A}&= \left[i2^{\frac{1}{4}}N^{IJ}\left( d\Psi +\tfrac{i}{2}(\Psi -%
\bar{\Psi})\mathcal{I}^{-1}d\mathcal{N}\right)\hspace{-3pt}{}_{J}\right] \wedge \Phi _{I} \notag\\
&\hspace{2em} +\left[-i2^{\frac{1}{4}}\frac{1}{(\bar{X}NX)}X^{I}\left( d\Psi +%
\tfrac{i}{2}(\Psi -\bar{\Psi})\mathcal{I}^{-1}d\mathcal{N}\right)\hspace{-3pt}{}_{I}\right]\wedge \bar{\Omega}+h.c.\  \;.
\label{F4decompalt}
\end{align}%
Substituting \eqref{eq:dPsi} one finds precisely the same expressions as \eqref{F4decomp} in the main text.


\begin{thebibliography}{99}
\bibitem{mohaupt1} V.~Cort\'es, C.~Mayer, T.~Mohaupt and F.~Saueressig, 
\textit{Special Geometry of Euclidean Supersymmetry I: vector multiplets},
JHEP \textbf{03} (2004) 028 [hep-th/0312001].

\bibitem{mohaupt2} V.~Cort\'es, C.~Mayer, T.~Mohaupt and F.~Saueressig, 
\textit{Special Geometry of Euclidean Supersymmetry II: hypermultiplets and the c-map}, 
JHEP \textbf{06} (2005) 024 [hep-th/0503094].

\bibitem{mohaupt3} V.~Cort\'es and T.~Mohaupt, 
\textit{Special Geometry of Euclidean Supersymmetry III: the local r-map, instantons and black holes},
JHEP \textbf{07} (2009) 066 [0905.2844].

\bibitem{witvan84} B.~de~Wit and A.~Van~Proeyen, 
\textit{Potentials and Symmetries of General Gauged N=2 Supergravity: Yang-Mills Models}, 
Nucl.\ Phys.\ \textbf{B245} (1984) 89, 
%
A. Strominger, 
\textit{Special Geometry}, 
Comm.\ Math.\ Phys.\ \textbf{133} (1990) 163.

\bibitem{onebigref} N.~Seiberg and E.~Witten, 
\textit{Monopole Condensation, And Confinement In N=2 Supersymmetric Yang-Mills Theory}, 
Nucl.\ Phys.\ \textbf{B426} (1994) 19 [hep-th/9407087]; 
%
N.~Seiberg and E.~Witten, 
\textit{Monopoles, Duality and Chiral Symmetry Breaking in N=2 Supersymmetric QCD}, 
Nucl.\ Phys.\  \textbf{B431} (1994) 484 [hep-th/9408099];
%
S.~Kachru and C.~Vafa, 
\textit{Exact Results for N=2 Compactifications of Heterotic Strings}, 
Nucl.\ Phys.\ \textbf{B450} (1995) 69;
%
S.~Kachru, A.~Klemm, W.~Lerche, P.~Mayr and C.~Vafa, 
\textit{Nonperturbative Results on the Point Particle Limit of N=2 Heterotic String Compactifications}, 
Nucl.\ Phys.\ \textbf{B459} (1996) 537 [hep-th/9508155].

\bibitem{blackdev} S.~Ferrara, R.~Kallosh and A.~Strominger, 
\textit{N=2 Extremal Black Holes}, 
Phys.\ Rev.\ \textbf{D52} (1995) 5412 [hep-th/9508072];
%
S.~Ferrara and R.~Kallosh, 
\textit{Supersymmetry and Attractors}, 
Phys.\ Rev.\ \textbf{D54} (1996) 1514 [hep-th/9602136];
%
K.~Behrndt, D.~L\"{u}st and W.~A.~Sabra, 
\textit{Stationary solutions of N=2 supergravity}, 
Nucl.\ Phys.\ \textbf{B510} (1998) 264 [hep-th/9705169];
%
A.~Strominger, 
\textit{Macroscopic Entropy of N=2 Extremal Black Holes}, 
Phys. Lett. \textbf{B383} (1996) 39 [hep-th/9602111];
%
G.~Lopes Cardoso, B.~de Wit and T.~Mohaupt, 
\textit{Corrections to macroscopic supersymmetric black-hole entropy}, 
Phys.\ Lett.\ \textbf{B451} (1999) 309 [hep-th/9812082];
%
G.~Lopes Cardoso, B.~de Wit, J.~Kappeli and T.~Mohaupt, 
\textit{Stationary BPS Solutions in N=2 Supergravity with  $R^{2}$-Interaction}, 
JHEP \textbf{12} (2000) 019 [hep-th/0009234].

\bibitem{gb} G.~W.~Gibbons, M.~B.~Green and M.~J.~Perry, 
\textit{Instantons and Seven-Branes in Type IIB Superstring Theory}, 
Phys.\ Lett.\ \textbf{B370} (1996) 37 [hep-th/9511080].

\bibitem{Zumino:1977yh}
B.~Zumino,
\textit{Euclidean Supersymmetry and the Many-Instanton Problem},
Phys.\ Lett.\ {\bf B69} (1977) 369.

\bibitem{GST} M.~Gunaydin, G.~Sierra and P.~K.~Townsend, 
\textit{The Geometry of N=2 Maxwell-Einstein Supergravity And Jordan Algebras}, 
Nucl.\ Phys.\ \textbf{B242} (1984) 244.

\bibitem{inst1} J.~B.~Gutowski and W.~A.~Sabra, 
\textit{Euclidean N=2 supergravity},
Phy.\ Lett.\ \textbf{B718} (2012) 610 [1209.2029]. 

\bibitem{inst2} J.~B.~Gutowski and W.~A.~Sabra, 
\textit{Para-complex geometry and gravitational instantons}, 
Class.\ Quant.\ Grav.\ \textbf{30} (2013) 195001 [1210.2332].

\bibitem{AlvarezGaume:1981hm} L.~Alvarez-Gaume and D.~Z.~Freedman,
\textit{Geometrical Structure and Ultraviolet Finiteness in the Supersymmetric Sigma Model}, 
Comm.\ Math.\ Phys.\  {\bf 80} (1981) 443.

\bibitem{Bagger:1983tt} J.~Bagger and E.~Witten,
\textit{Matter Couplings in N=2 Supergravity},
Nucl.\ Phys.\ {\bf B222} (1983) 1.

\bibitem{Ferrara:1989ik} S.~Ferrara and S.~Sabharwal, 
\textit{Quaternionic Manifolds for Type II Superstring Vacua of Calabi-Yau Spaces}, 
Nucl.\ Phys.\ \textbf{B332} (1990) 317.

\bibitem{Vaughan:2012} O.~Vaughan, 
\textit{The r-map, c-map and black hole solutions}, 
PhD Thesis (2012).

\bibitem{Cortes:2015} V.~Cort\'es, P.~Dempster, T.~Mohaupt and O.~Vaughan 
\textit{Special Geometry of Euclidean Supersymmetry IV: the local c-map},
JHEP {\bf 1510} (2015) 066 [1507.04620].

\bibitem{Hull} C.~M.~Hull, 
\textit{Duality and the signature of space-time}, 
JHEP \textbf{11} (1998) 017 [hep-th/9807127].

\bibitem{Hull:1998br}
  C.~M.~Hull and B.~Julia,
  \textit{Duality and moduli spaces for timelike reductions,}
  Nucl.\ Phys.\ {\bf B534} (1998) 250 [hep-th/9803239].
  
  \bibitem{Cremmer:1998em}
  E.~Cremmer, I.~V.~Lavrinenko, H.~L\"u, C.~N.~Pope, K.~S.~Stelle and T.~A.~Tran,
  \textit{Euclidean signature supergravities, dualities and instantons,}
  Nucl.\ Phys.\ {\bf B534} (1998) 40
  [hep-th/9803259].

\bibitem{Bodner} M.~Bodner, A.~C.~Cadavid and S.~Ferrara, 
\textit{(2,2) vacuum configurations for type IIA superstrings: N=2 supergravity Lagrangians and algebraic geometry}, 
Class.\ Quant.\ Grav.\ \textbf{8} (1991) 789.

\bibitem{cremmer} E.~Cremmer, B.~Julia and J.~Scherk, 
\textit{Supergravity in theory in 11 dimensions}, 
Phys.\ Lett.\ \textbf{B76 }(1978) 409.

\bibitem{Ferrara:1988ff} S.~Ferrara and S.~Sabharwal, 
\textit{Dimensional Reduction of Type II Superstrings}, 
Class.\ Quant.\ Grav.\ \textbf{6} (1989) L77.

\bibitem{Bergshoeff:2007cg} E.~A.~Bergshoeff, J.~Hartong, A.~Ploegh, 
J.~Rosseel and D.~Van den Bleeken, 
\textit{Pseudo-supersymmetry and a tale of alternate realities}, 
JHEP {\bf 07} (2007) 067 [0704.3559].

\bibitem{Ferrara:1990dp} S.~Ferrara, M.~Bodner and A.~C.~Cadavid, 
\textit{Calabi-Yau supermoduli space, field strength duality and mirror manifolds},
Phys.\ Lett.\ \textbf{B247} (1990) 25.

\bibitem{Candelas} P.~Candelas and X.~de la Ossa, 
\textit{Moduli Space of Calabi-Yau Manifolds},
Nucl.\ Phys.\ \textbf{B355} (1991) 455.

\bibitem{Fre:1995bc} P.~Fre and P.~Soriani,
\textit{The N=2 wonderland: From Calabi-Yau manifolds to topological field theories},
Singapore, Singapore: World Scientific (1995) 468p.

\bibitem{Green:1987mn} M.~B.~Green, J.~H.~Schwarz and E.~Witten,
\textit{Superstring Theory. Vol. 2: Loop Amplitudes, Anomalies And 
Phenomenology,} Cambridge, UK: Univ.\ Pr.\ (1987) 596p (Cambridge 
Monographs On Mathematical Physics).

\bibitem{Louis:2002ny} J.~Louis and A.~Micu, 
\textit{Type 2 theories compactified on Calabi-Yau threefolds in the presence of background fluxes},
Nucl.\ Phys.\ {\bf B635} (2002) 395 [hep-th/0202168].

\bibitem{Hull:1998vg}
  C.~M.~Hull,
  \textit{Timelike T duality, de Sitter space, large N gauge theories and topological field theory,}
  JHEP {\bf 07} (1998) 021[hep-th/9806146].
\end{thebibliography}
\end{document}